\documentclass{article}

\usepackage[T1]{fontenc}
\usepackage{amsmath}
\usepackage{graphicx}
\usepackage{gensymb}
\usepackage{titlesec}
\usepackage{lipsum}
\usepackage{cite}

 \titleformat{\section}
  {\normalfont\scshape}{\thesection}{1em}{}
    
\begin{document}
\title{Quantum method for computations of flows around a body with Data analysis}
\author{
  Zhao, Maomao\\
  \textit{University of Toronto}\\
  \textit{maomao.zhao@mail.utoronto.ca}
  \and
  Liu, Yufei
 }

\maketitle

\begin{abstract}
In this paper we briefly introduce the quantum methods for computations of the drag coefficients for flows around a body, using the flows around a rigid sphere as an example, and we aim for comparing the wake under quantized environment and the classical computational strategies for finding the drag force. This paper however doesn't provide discussion on the pressure distribution over the surfaces of a spherical body and neither the analytical solution for the flows around an airfoil due to the greatly limited computational capacities and resources. 
\end{abstract}

\section*{Preface}
The fundamental problem of aerodynamics is to find the action force allocated along with the surface of a body in the flow (such as the lift and drag force on airfoils, etc), thus we shall first study how the physical quantities distribute in the entire flow field. The relevant theory development peaked at the time shortly after Joukowski and Prandtl dedicated their effort. However Navier-Stoke equation and Reynolds equation had become to their fellow researchers formidable mathematical obstacles ahead of them. In order to attain accurate data records in a flow field, we have to be aware of the complete figure of the field with great details, such as the regulatory development routine for the boundary layer, the laminar-turbulent transition, the development of the separation region, and the mechanism of turbulence energy dissipation in all sections. Only through unveiling  the mysteries of these principles, can we learn more about the pressure distribution as well as the appropriate circumstances for applying Euler equations. When we study turbulence into depth, we shall refer to a few people of importance: G.I Taylor, L. Prandtl, Von Karman, Kolmogorov, and R. Kraichnann. It's worthy to note that,  while Potential flow theory and Viscous flow theory can neither provide complete sustainable explanations for the physical principles behind the development of a flow field, the quantum flow field we discuss in this paper shall be recognized as a hypothesis. The studies we are doing today, apart from what our elder generation of researchers had aimed for, are to actualize the quantization of flow fields by incorporating a generalized de Broglie relation.

\section*{Part A \textit{Framework}}
The diagram of the quantized flow field is as shown in \textit{Figure 1}.

\begin{figure}
  \includegraphics[width=\linewidth]{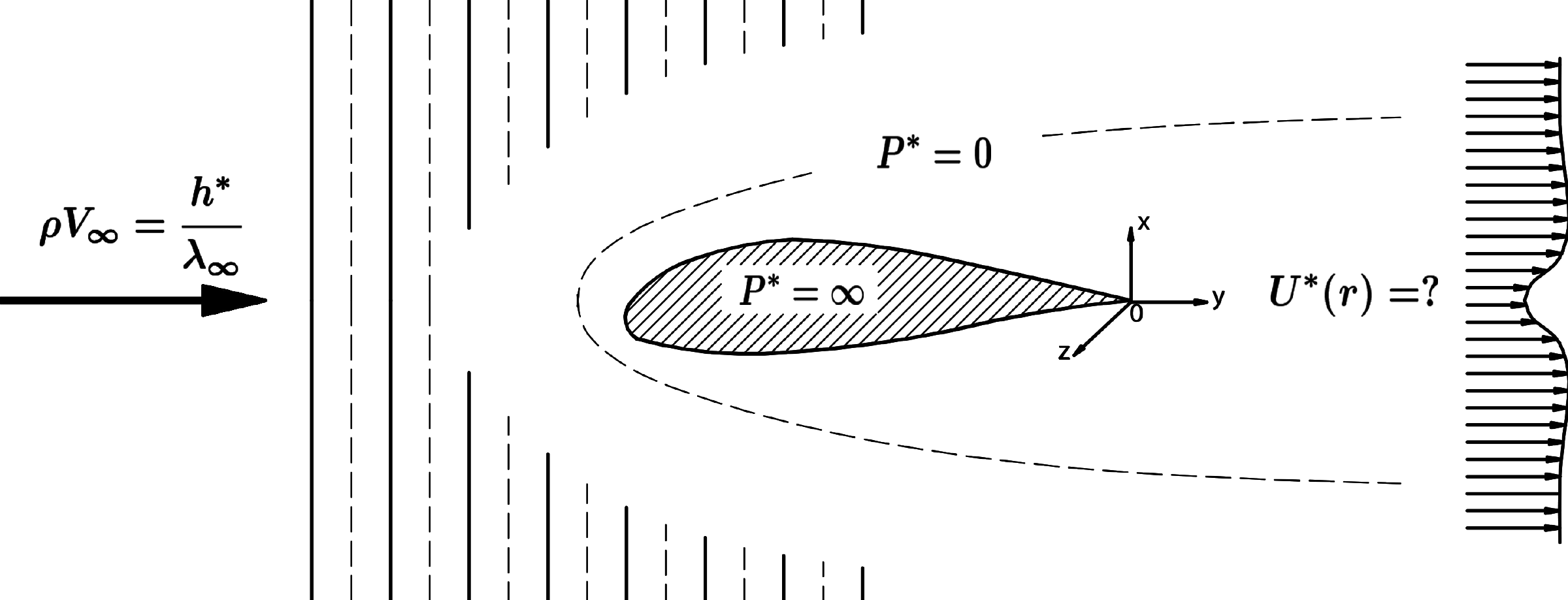}
  \caption{\textit{The depiction for quantizing the flow field with a rigid body (airfoil shape in this case)}}
    \label{fig:fig1}
\end{figure}

The topic regarding traditional flow field problems would focus on the physical quantity distributions of a uniform straight flow along with the surface of a rigid body in an arbitrary shape, and on the computation of the resulting force distributions acting on the particular object. 
Similarly to the classical potential theory where we use a dipole instead of a sphere, or use the distribution function of a source/sink instead of an ordinary airfoil model, in the quantization theory of a flow field, we replace the uniform straight flow with a plane wave and the body in the field with a generalized potential function. But we ought to mention that here the plane wave, is in fact a generalized de Broglie wave, i.e. a material wave; is not an oscillation from any classical physics phenomena. Once the geometrical figuration of an object in the flow field is steady to the observers in classical potential theory, there exists an exact potential function, which is the potential of velocity, and this potential function will be the solution of the entire flow field. In resemblance of the potential theory, the quantization method of a flow field thus seeks a particular wave function for some generalized Schrodinger equation.

Mathematically speaking, the generalization of de Broglie relation doesn't encounter any difficulties in the effort doing logical advancement. 
The following equations hold up the fundamental structures in the quantization of a flow field:

1.generalized de Broglie relation:
\begin{equation} 
\rho V_{\infty}= \frac{h^{\ast}} {\lambda_{\infty}} ~~or~~ \rho \overrightarrow{V_{\infty}}=\hbar^{\ast}\overrightarrow{k} 
\end{equation} 
\begin{equation}
E_{\infty}=h^{\ast}\nu 
\end{equation}

2.generalized Schrodinger equation:
\begin{equation}
\left[\frac{-(\hbar^{\ast})^2}{2 \rho}\bigtriangledown^2+P(x,y,z)\right]\varphi=E_{\infty}\varphi
\end{equation}

3.generalized velocity(in probability) distributions:
\begin{equation}
U(x,y,z)=\frac{\hbar^{\ast}}{2i\rho}\left[\varphi^{\ast}\bigtriangledown\varphi-\varphi\bigtriangledown\varphi^{\ast}\right]
\end{equation}

The above equations are the theoretical framework of a flow over a body under certain circumstances completely made up of material waves, as already shown in \textit{Figure 1}. As indicated, $\rho V_{\infty}$ and $E_{\infty}$ are respectively the quantity of the momentum and of the kinetic energy for the uniform straight flow in unit volume. And $E_{\infty}=\frac{1}{2}\rho V_{\infty}$, where $h^{\ast}$ is the generalized Planck constant, its according dimension is the fluid angular moment in the unit volume. $\lambda_{\infty}$ is a generalized de Broglie wavelength; $k= \frac {2\pi} {\lambda_{\infty}}$, $\hbar^{\ast}=\frac{h^{\ast}}{2\pi}$, $P(x,y,z)$ is the generalized potential function; its dimension corresponds to the energy in unit volume; similarly, the dimension for $U(x,y,z)$ stands for the velocity.
If we plug in the generalized de Broglie relation (1) into the explicit expression for Reynolds Number:
$$Re=\frac{\rho V_{\infty} L}{\mu}$$
We can get the most significant dimensionless quantity in our wave features, i.e. the generalized Re number:
$${Re}^{\ast}=\frac{h^\ast}{\mu}\left(\frac{L}{\lambda_{\infty}}\right)$$

These are the \textit{similarity criteria} for dynamics in a quantized environment, and in the above formula, both $h^{\ast}$ and $\mu$ are attached to various numerical values with respect to different fluid mediums; in addition, they will have the same dimension. The existing and only admitted dimensionless parameter for all types of waves diffractions exhibits as $\frac{L}{\lambda_{\infty}}$. The ratio for $\frac{h^{\ast}}{\mu}$ settles the exact correspondence in coordinate parametrization between $Re^{\ast}$ and $Re$.

\section*{Part B \textit{Model Interpretations}}
\begin{figure}
  \includegraphics[width=\linewidth]{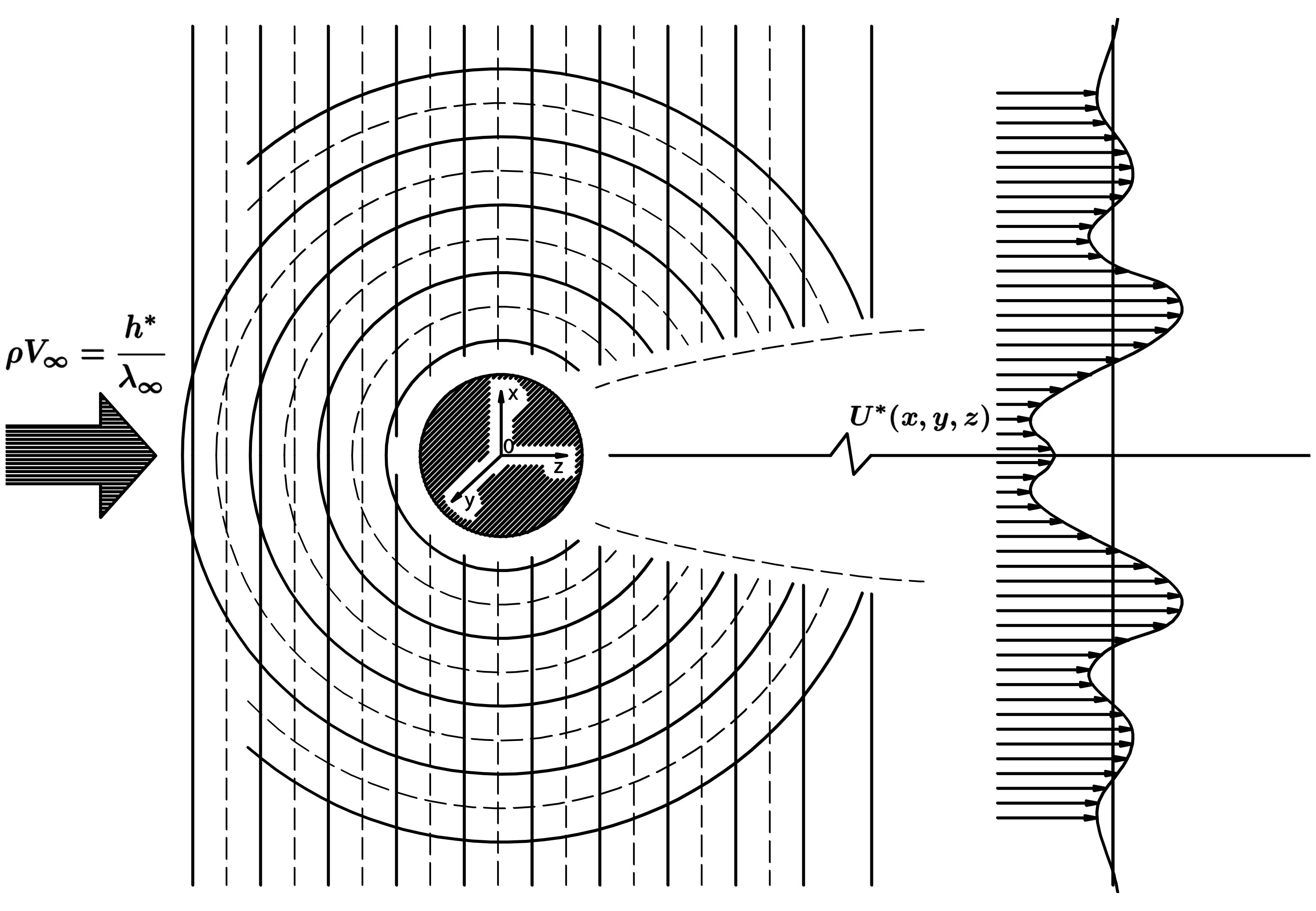}
  \caption{\textit{Quantized flow field and the formation of the velocity distribution in probability.}}
    \label{fig:fig2}
\end{figure}
As shown in Figure 2, when we set the origin for polar coordinates to be the center of a sphere, the generalized potential function $P(x,y,z)$ would have a simplified expression written in polar coordinates:
\begin{equation}
P=P(r)
\end{equation}

and \begin{equation*}
   P=
   \begin{cases}
      0, & \text{when}\ r > a\\
      \infty, & \text{when}\ r < a
   \end{cases}
\end{equation*} where $a$ is the radius of the spherical body.
Here 
\begin{equation}
\varphi = 0 |_{r=a}
\end{equation} is as required for the boundary condition.

The formulas (5) and (6) are the \textit{no slipping condition} and \textit{impenetrability condition} on the surface of the body in a quantized flow field. Noticing that the problem of the flow over a sphere, as we have discussed from above, has turned into a spherical symmetric generalized potential function acting on the plane wave presenting the uniform straight flow as a scattering process, and such a problem can have a relatively simple solution with partial wave method. At this moment the generalized Schrodinger equation may have an asymptotic solution:
\begin{equation}
\varphi(r, \theta) ~\overrightarrow{\scriptscriptstyle{r\rightarrow \infty}}~e^{i k z} + f(\theta) \frac{e^{i k r}}{kr}\end{equation}

where \begin{equation}
f(\theta)=\sum^{\infty}_{l=0} (2l+1)~e^{i\delta_l}~sin~ \delta_l ~P_l(cos \theta)
\end{equation}
 which is the so called scattering amplitude, and $|f(\theta)|^2$ is the differential cross section. $l = 0, 1, 2, \ldots$ correspond to the partial waves $s, p, d,$ etc. $\delta_l$ is the phase shift for the $lth$ partial wave. And $P_l$ is the $lth$ term for legendre polynomial.

\section*{Part C \textit{Data Analysis}}
Substituting (7) into (4), we can obtain the asymptotic solution for the velocity distribution in probability, i.e the solution to the far-field wake behind the spherical body. Because of our limitation on the computational skills and the enormous amount of work required for further analysis, here we only discuss the velocity distribution in probability parallel to the z-axis for some cross-section of a wake. Its asymptotic solution has the form:
\begin{equation}
U = \frac{\hbar k}{\rho} \left[ 1 + |f(\theta)|^2~\frac{cos~ \theta} {kz} + (1+cos~ \theta)~\frac{f_a~ cos(KR) - f_b~ sin(KR)}{kz}\right]
\end{equation} where $z= r~cos~ \theta~=~30~\pi~a$, $f_a$ and $f_b$ are the real and imaginary parts for $f(\theta)$.
This is an approximated one omitting some higher order terms when $r>>a$, \textit{Figure 3} displays it in its typical prominence with 2 examples of $ka = 1.6$ (as $\theta$ grows to $80\degree $) and $ka = 10$ (as $\theta$ grows to $50\degree$.)

\begin{figure}
  \includegraphics[width=\linewidth]{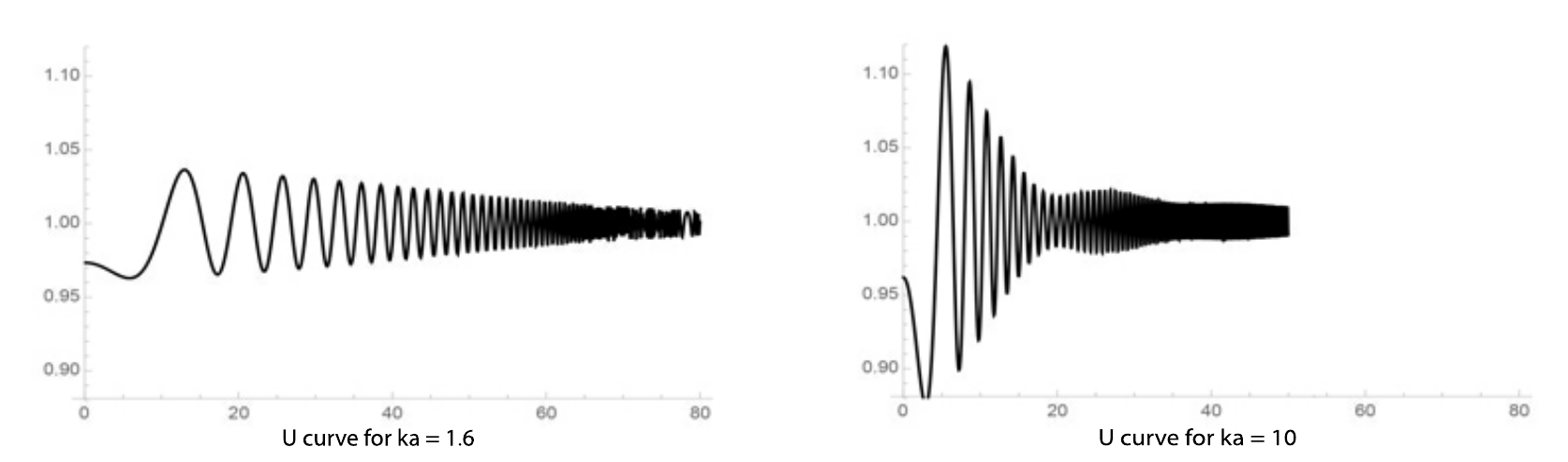}
  \caption{\textit{Typical U curves.}}
    \label{fig:fig3}
\end{figure}

In a graph carrying the main properties of the curve $U$, it demonstrates which we shall consider as the basic form of what $U$ really is, the velocity distribution in terms of its probability natures, and thus we shall  average $U$ in some way to get a result assimilating the curves being confirmed from wake attending experiments. But it's worth noting that even if we don't take some form of average to the $U$ curve, there still always exists a \textit{deficit area} on the graph for the velocity near z-axis (when $\theta \sim 0\degree$). When $ka$ becomes big enough, we can see a special character called \textit{modulation} in the "tail" (i.e. perpendicular to the z-axis)  of the curve $U$ describing the velocity in a quantized wake. The \textit{"modulation"} implies that when $ka$ ($\frac{\pi*L}{\lambda_{\infty}}$) increases, wake width decreases, and it may also suggest there exists a multitude of \textit{wave packets} in the flow field but outside of the wake. We call the indented parts between these wave packets \textit{wasp-waist shape} (when $ka$ is small, we can't see a second wave packet on the $U$ curve so we may consider the \textit{transient point} to be the \textit{wasp-waist shape} as the first wave packet narrowing down until where its two horizontal "boundary lines" become relatively parallel.) These mysterious figures of the velocity distribution in probability are first found in data analysis, which shall draw our special attention since they may be encrypted with unpredictable physical importance. We can at least propose that one among many of the quantum flow field properties is the complex interference between the bouquets of de Broglie waves. In classical flow field theory, the transition mechanism turning from turbulence flow to laminar flow and the displacement of such a transition lack a creditable theoretical explanation, and therefore this becomes one of the key reasons why we are still having troubles determining the numerical values of the dynamical parameters (especially for the lift and drag force parameters) on bodies in a flow field. However, if we use the quantization method of the flow over a body, including formulas of computations in the entire flow field and even for the computations of the mechanism of emerging turbulence flow, we may lead ourselves to an overall new prospect for analysis of the flow field. 

In fact, 
\begin{equation}
\psi(p)~\propto \int{\varphi(r)~e^{-\frac{i}{\hbar} \cdotp \overrightarrow{p}\cdotp \overrightarrow{r}}}~dr
\end{equation}
i.e. the momentum distribution in probability can be extracted from the Fourier transform of wave functions to the generalized Schrodinger equation, which implies we may get the spectral function directly. This is already defined by de Broglie relation $\overrightarrow{p}=h^{\ast} \overrightarrow{k}$. And it also constitutes the basics of the statistical description of turbulence, i.e. the concepts behind the energy spectral density function must come from the wave function in the momentum space, which is in resemblance with the turbulence spectral theory 
$${U'}^2=\int{\epsilon(n)} ~dn$$
forwarded by G.I.Taylor. 

In contrast to what G.I.Taylor believed about the mechanism of energy-spectral analysis, we think that it's more reasonable and perceivable to consider the family of waves in momentum space to be the generalized de Broglie waves, which is a physics cause in principle of the energy spectral density function. By the observation of some phenomenon in data analysis, we accommodate the U function into some form of its average and thus lead to the formula of the computation for drag under quantized environment:
$$C_{\scriptstyle{D}}=\eta\int{\left(1-\frac{\overline{U}}{V_{\infty}}\right)}~ds + \varepsilon\left(ka\right)$$
Where $\eta$ is an undetermined constant, $U, ~\overline{U}$ is the velocity distribution in probability and its average (in quantum flow field), which is very close to the classical integral computational method for drag. $\varepsilon\left(ka\right)$ stands for some necessary complementing term added to the drag formula upon the fact that the occurring pattern in the graph of velocity distribution in probability inside a wake is in correspondence to the energy of turbulence fluctuation. But mind that it differs from the energy dissipation term of the turbulence theory. Given that we already have the simplified expression of a solution to the velocity in probability of flow over a sphere (as shown in equation (9)), taking account of the existence of the \textit{wasp-waist shape} between wave packets, we shall expect that $\varepsilon\left(ka\right)$ having a mathematical form of:
$$\varepsilon \left(ka\right)=\int^{\infty}_{-\infty}{\alpha e^{-\beta x^2}}~dx$$
Where $\alpha,~\beta$ are parameters involving contents from the \textit{wasp-waist shapes,} which shall be chosen based on individual experiments using simulations.

In our data analysis by simulations, we emphasize some of the features which should be carefully studied in the future:

1)	when $ka$ approaches 0, the velocity distribution in probability would be back to the classical velocity profile as for $Re \rightarrow 0$, i.e. the width of the wake grows to infinity (or similarly $C_D \rightarrow \infty$.) This shows for $ka \rightarrow 0$ or for $Re \rightarrow 0$ the quantized flow field and the classical flow field would lead to the same solution for the velocity distribution. \textit{Figure 4} shows the $U$ curves for $ka = 0.000001,~0.01,~0.1,~0.9,~1.6,~2.3$ with each curve rotated $90\degree$ counter-clockwise.

\begin{figure}
  \includegraphics[width=\linewidth]{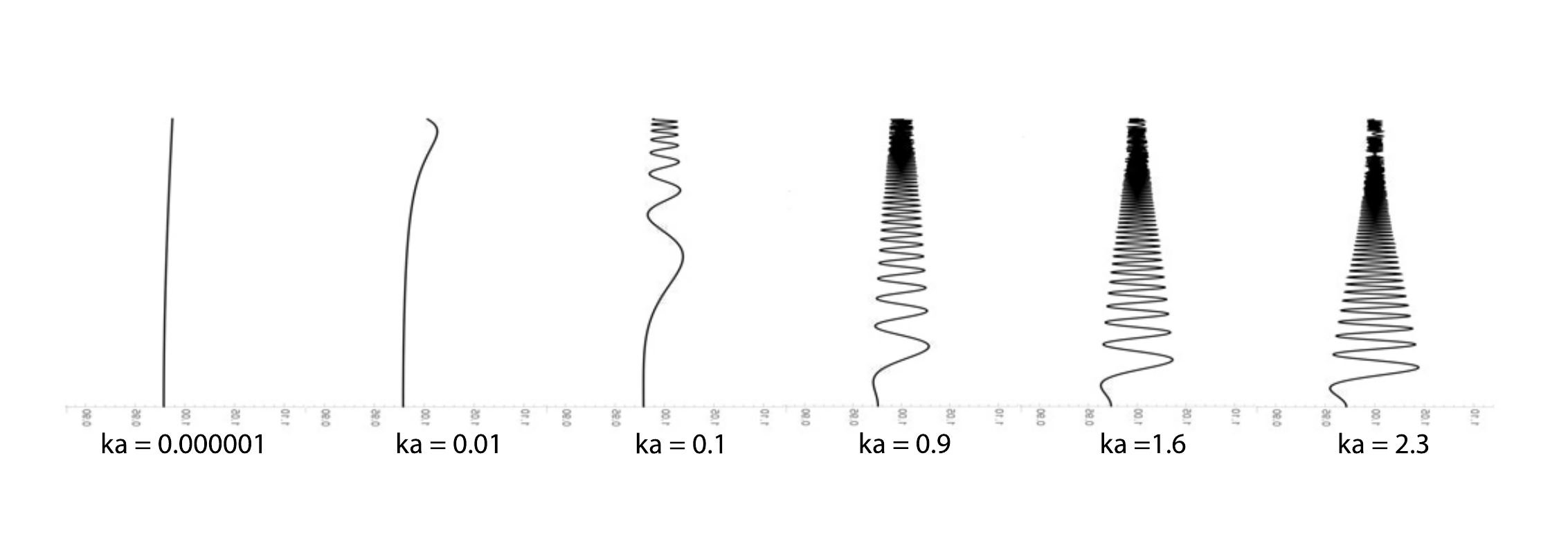}
  \caption{\textit{U curves flipped $90\degree$ counter-clockwise for various $ka$ values}}
    \label{fig:fig4}
\end{figure}

2)	If we pick out the first \textit{wasp-waist shapes} from each $U$ graph and place them all together into one curve as \textit{Figure 5} displays (for 17 typical $ka$ values: 0.000001, 0.01, 0.1, 0.9, 1.6, 2.3, 3, 4, 5, 6, 7, 8, 9, 10, 11, 12, 13), we first see a drastic drop off between $0.01<ka<0.1$ since we can't even see clear oscillating features especially when $ka$ becomes smaller than 0.01. So we may assert that the first \textit{wasp-waist shape} appears very close to $90\degree$, in other words, close to \textit{infinity} if we use cartesian coordinates. Moreover the rate of change of the positions where the first wasp-waist shape figures appearing on $U$ curves getting closer to $\theta = 0$ relatively accelerates for another time around the region $1.8<ka<3.5$. We hold that this is a primary reason for the existence of the drag crisis. If we leave the "red curve" on \textit{Figure 5} only and adjust \textit{ka values} to correct scales, i.e. $x=30\pi \tan(\theta)$, we will get a curve as depicted in the graph of \textit{Figure 6}.  But at this point our knowledge about the $wasp~waist~shape$ as well as where it would appear is minimal. So we should note that our reading to the locations of the first $wasp~waist~shapes$ now is highly depending on visual premises, although our assumptions as follows would maintain some theoretical validity, this evaluation should only be considered as a hypothetical reference. Notice the first wave packet doesn't emerge distinctively until $ka$ reaches $0.9$, and the second wave packet appears when $ka>3.8$, so for $0.9<ka<3.8$ we claim that the \textit{wasp-waist shape} sits on the region separating the first wave packet from the "tail", i.e. the part where the "amplitude" of the wave increases and the part where the amplitude stays relatively constant; and for $ka<0.9$ (where the growth rate of the wave amplitude isn't quite distinguishable), we pick our point based on the same assumption as before but this reading may include some inconsistency needed to be clarified in future studies because of the limited information we know about the nature behind these wave packets and time required for computations in depth supporting the legitimate evidence. 

\textit{(Note that in Figure 5 and 6 we only compute the U values from $\theta = 0$ to $\theta = 50\degree$ for $ka$ is greater than $5$ whereas $0$ to $80\degree$ for $ka \leq 5$ and $0$ to $85\degree$ for $ka=0.3,~0.5,~0.7,~0.9,~1.3$ due to the exponentially increasing number of points we need to select for computations as $ka$ grows bigger.)}

In quantum flow field, $Re_{critical}$ would attain a wide range of efficiency. This is related to the fact that the impact of S wave (the component wave corresponding to the first term in the series embedded in the simplified solution of (8)) on the cross-section of the scattering reduces to a much less significant extent, and causing by generalized Ramsauer effect, reduces to 0 when $ka$ reaches $\pi$. The generalized Ramsauer effect depicts that every component wave would encounter a \textit{transmission} when $ka$ equals certain values. Then the cross-section would become smaller as a result of a transmission. Thus this overall affects the value of the drag coefficient $C_D$.

\begin{figure}
  \includegraphics[width=\linewidth]{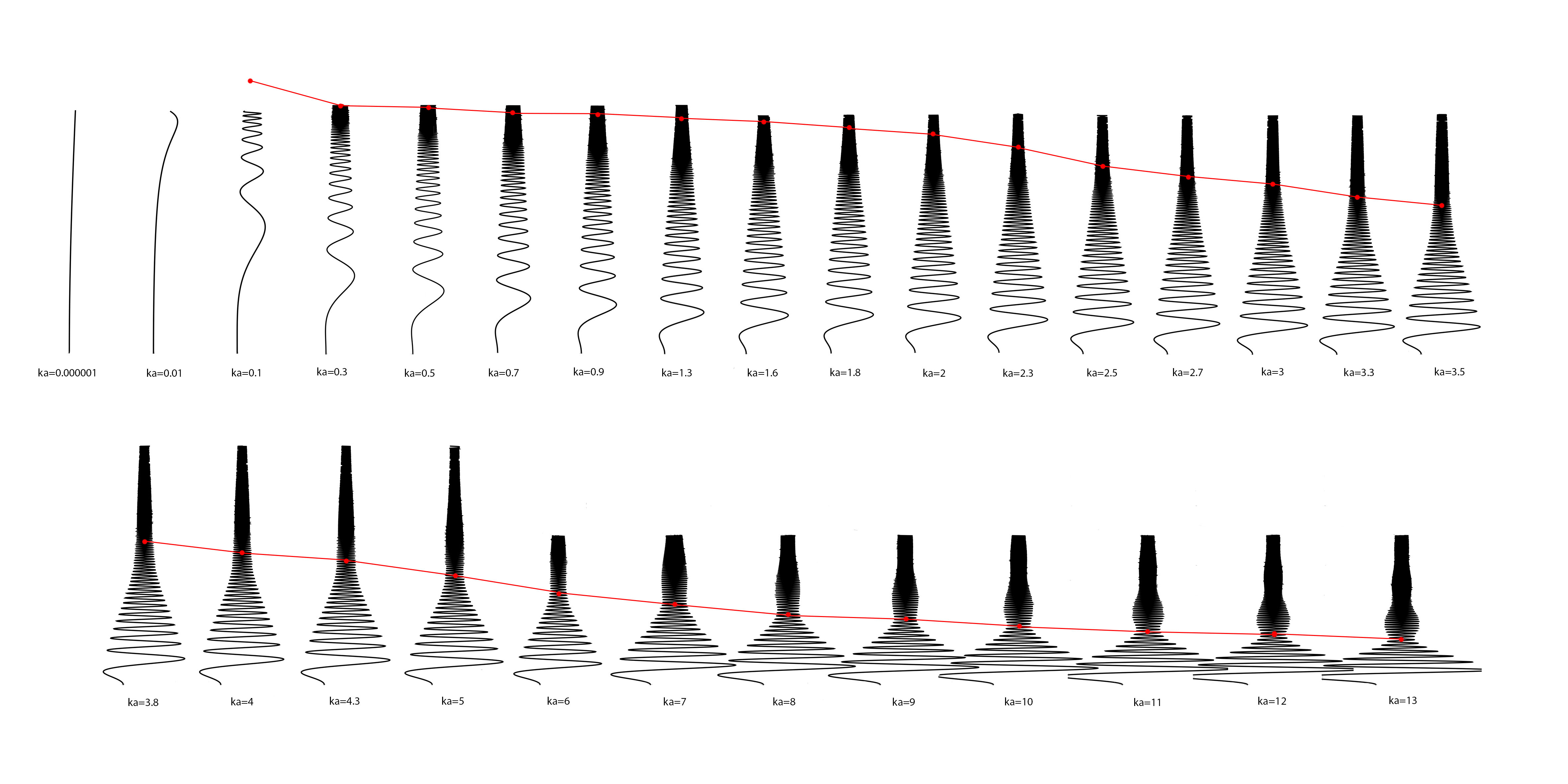}
  \caption{\textit{Red line linking each wasp waist shapes and the shadowing triangles.}}
    \label{fig:fig5}
\end{figure}

\begin{figure}
  \includegraphics[width=\linewidth]{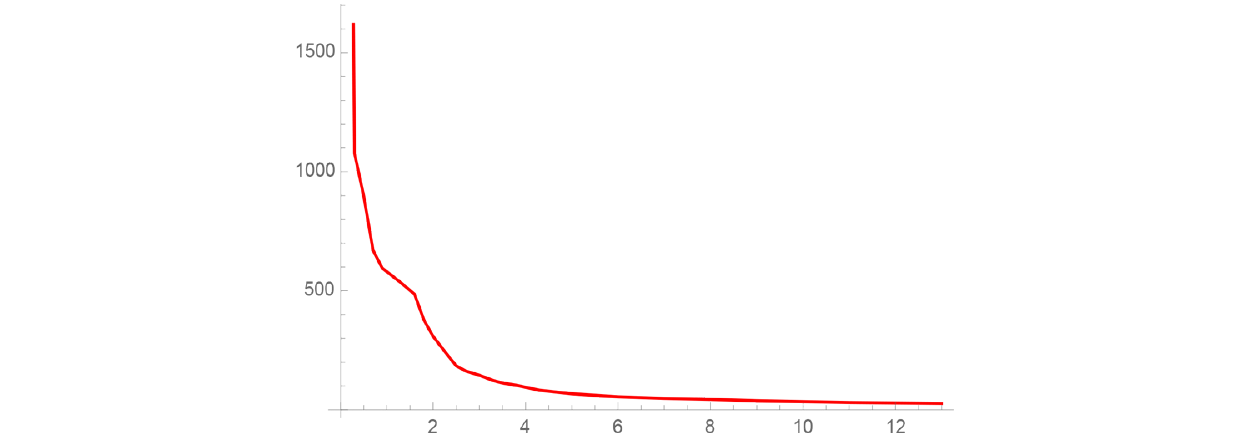}
  \caption{\textit{The curves for the positions of the first wasp waist shapes}}
    \label{fig:fig6}
\end{figure}

Notice that the \textit{wasp-waist shape} on each $U$ curve is not unique, there exist multiple ones especially when $ka$ becomes greater than 10 (in our original calculation when $ka$ grows up to 40 we saw as many as 4 of these shapes on each $U$ curve.) This is a result emerging specifically in wave mechanics. The quantization method transforms the classical flow field into a capricious interfering wave family where the existence for modulation is a primary characteristic.  From the figure we shall notice the undeniable importance of the first \textit{wasp waist shape} to the calculation of drag coefficient; thus it would be reasonable to predict the rest multiple wasp waist shapes on the $U$ curves when $ka$ gets higher also devote into the formation of drag coefficients, i.e. in last integral formula presented above the $\varepsilon \left(ka\right)$ term is an essential term directly related to the number or the properties of all the wasp waist shapes. But at the moment we could not provide precise information about them or about the numerical relations between the other multiple wasp waist shapes and the drag coefficient. In the future potential researches following this paper, we encourage people to specifically work on these features.

\section*{Remarks on the Boundary layer theory by Prandtl}

Boundary Layer Theory by Prandtl is the most widely accepted theory for illustrating the actual process of flow over a body to our current researchers in the field of aerodynamics. Yet we still lack some valid theoretical explanations for experimental data, i.e. when $Re$ number grows greater than $Re_{critical}$, why are both drag coefficient decreasing and its increasing gradual changes over the entire phase instead of sudden shifts? Prandtl's well known experiment was to place a thin metal ring to the windward side of a sphere and to make the separation point on the surface of that sphere move to the lower course; it doesn't provide coordinate-dependent demonstration especially on the time-space evolution for this separation process. On the contrary, suppose the transition from laminar flow to turbulence flow is a morphological abruptness, then there will be a great scale of topological changes, and $Re_{critical}$ is a relatively small region, further, $C_D$ will have a relatively minor change. However this doesn't agree with experimental facts. Not only $C_D$ changes over a wide range of efficiency, but its numerical value grows back slowly as $Re$ becomes greater than $5 \sim 6 \times 10^5$. And the separation point on the surface of the sphere moves to the upper course instead. Boundary Layer Theory has so far failed to provide valuable insights on this disagreement.

\section*{Acknowledgements}
The authors are sincerely grateful to the people for making this paper possible including family and friends. 

The following names shall be mentioned for their invaluable contributions to this paper as mentors and friends: 

The late professors Zhuang Fenggan, and Xu Huafang, founders of China's studies in Aerodynamics who had encouraged the authors for the development of this theory; Professor Li Chunxuan at Beihang University who offered treasured consultation for advancing this paper; Professors Zhang Fuqing, Liu Ziqiang and Yuan Youxin for their detailed advice during the discussions about this paper; Mrs.Tan Liuning for her generous help financially and her first approach in some preliminary computations which turned out to be significant results.

Pratyush Sarkar, Ethan Yale Jaffe, Freid Tong, Michael Yu, who all have been great friends and been utterly patient providing the author professional opinions regarding the paper and other mathematical topics; Yuqing Tang, Tom Wang, and Guanchu Liu for their help initiating this paper; Li, Andrew, Tomas, Zhifei, Eric, Yuan, David, Christopher, Adam, Samer, Jonathan, and all the other friends without whom this paper cannot be possible.

The authors also feel inexpressibly thankful to Department of Mathematics, University of Toronto and all the faculty members and students for making the department a lovely place to learn knowledge.

  Formulas (4) to (8) \cite{Schiff} have the almost identical presentations to the similar setting ups for the quantum scattering theory. The referenced books are the ones we used when consulting background knowledge from quantum mechanics \cite{Mott} \cite{Bohm}, but we believe most of the earlier books studying quantum should all have these formulas and it's hard to track the true origin of them. The theoretical as well as the computational results presented in this paper don't have relative fellow researchers' work to be compared with, and the theory been developed behind this paper originated from the author Yufei Liu's book \textit{An introduction to Quantum Methods for Flows around a Body} published in 2013 (currently only formally available in Mandarin.) As almost all of the named theories or problems in the text aside from the new method introduced the first time are very well known and should not cause any confusion, we decided not to list any specific articles or books for references aside from the above ones.

\end{document}